\documentclass{article}
\usepackage{hyperref}
\usepackage{arxiv}

\usepackage[utf8]{inputenc} 
\usepackage[T1]{fontenc}    
\usepackage{mathrsfs,amsmath,amssymb,bm}
\usepackage{lipsum}
\usepackage{amsfonts}
\usepackage{epstopdf}
\usepackage{makecell, rotating}
\usepackage{multirow}
\usepackage[nottoc]{tocbibind}
\usepackage{array}
\usepackage{listings}
\usepackage{cleveref}
\usepackage{caption}
\usepackage{authblk}
\usepackage[utf8]{inputenc} 
\usepackage[T1]{fontenc}    
\usepackage{url}            
\usepackage{booktabs}       
\usepackage{nicefrac}       
\usepackage{microtype}      
\usepackage{xcolor}         
\usepackage{multirow}
\usepackage{graphicx,subfigure}
\usepackage[ruled,linesnumbered]{algorithm2e}
\usepackage{tabularx}

\usepackage{gensymb}
\usepackage{float}
\usepackage{mwe}

\numberwithin{equation}{section}

\title{A multi-category inverse design neural network and its application to diblock copolymers}

\author[1,$\ddagger$]{Dan Wei}
\author[1,$\ddagger$]{Tiejun Zhou}
\author[1]{Yunqing Huang}
\author[1]{Kai Jiang\thanks{\texttt{kaijiang@xtu.edu.cn}}}

\affil[1]{Hunan Key Laboratory for Computation and Simulation in Science and Engineering, Key Laboratory of Intelligent Computing and Information Processing of Ministry of Education, School of Mathematics and Computational Science, Xiangtan University, Xiangtan, Hunan, China, 411105.}
\affil[$\ddagger$]{These authors contributed equally to this work.}

\renewcommand{\headeright}{}
\renewcommand{\undertitle}{}

\begin{document}
\maketitle

\begin{abstract}
  In this work, we design a multi-category inverse design neural network to map ordered periodic structure to physical parameters. The neural network model consists of two parts, a classifier and Structure-Parameter-Mapping (SPM) subnets. The classifier is used to identify structure, and the SPM subnets are used to predict physical parameters for desired structures. We also present an extensible reciprocal-space data augmentation method to guarantee the rotation and translation invariant of periodic structures.
    We apply the proposed network model and data augmentation method to two-dimensional diblock copolymers based on the Landau-Brazovskii model.
	Results show that the multi-category inverse design neural network is high accuracy in predicting physical parameters for desired structures. Moreover, the idea of multi-categorization can also be extended to other inverse design problems.
\end{abstract} 
	

	\section{Introduction}\label{sec:intrd}
	
	Material properties are mainly determined by microscopic structures. Therefore, to obtain satisfactory properties, how to find desired structures is very important in material design. The formation of ordered structures directly relies on physical conditions, such as temperature, pressure, molecular components, geometry confinement. However, the relationship between ordered structures and physical conditions is extremely complicated and diversified.  
	A traditional approach is a trial-and-error manner, i.e., passively finding ordered structures for given physical conditions. 
	This approach, in terms of solving direct problem, is time-consuming and expensive.
	A wise way is inverse design that turns to find physical conditions for desired structures.  
	
	In this work, we are concerned about the theoretical development of inverse design method for block copolymers. Block copolymer systems are important materials in industrial applications since they can self-assemble into innumerous ordered structures. There are many solving direct problem approaches of block copolymer systems, such as the first principle calculation \cite{beu2005first}, Monte Carlo simulation \cite{he2001self,sugimura2021monte}, molecular dynamic \cite{lemak2003molecular}, dissipative particle dynamics \cite{ortiz2005dissipative, gavrilov2013phase}, self consistent field simulation \cite{fredrickson2006equilibrium}, and density functional theory \cite{fraaije1993dynamic}.
	In the past decades, a directed self-assembly (DSA) method has been developed to inverse design block copolymers. 
	Liu et al.~\cite{liu2010integration} presented an integration scheme of block copolymers directed assembly with 193 nm immersion lithography, and provided a pattern quality that was comparable with existing double patterning techniques. Suh et al.~\cite{suh2017sub} obtained nanopatterns via DSA of block copolymer films with a vapour-phase deposited topcoat. 
	Many DSA strategies have been also developed for the fabrication of ordered square patterns to satisfy the demand for lithography in semiconductors \cite{li2021square, ouk2003epitaxial, ji2011three, chuang2009templated}.
	
	With the rise of data science and machine learning, many deep-learning inverse design methods have been developed to learn the mapping between structures and physical parameters \cite{malkiel2017deep,gahlmann2022deep,liu2018training,john2017nanophotonic}. These new techniques and methods are beginning to study block copolymers. Yao et al. combined machine learning with  self consistent field theory (SCFT) to accelerate the exploration of parameter space for block copolymers \cite{yao2021deep}. Lin and Yu designed a deep learning solver inspired by physical-informed neural networks to tackle the inverse discovery of the interaction parameter and the embedded chemical potential fields for an observed structure \cite{lin2022Deep}. Based on the idea of classifying first and fitting later, Katsumi et al. estimated Flory-Huggins interaction parameters of diblock copolymers from cross-sectional images of phase-separated structures \cite{hagita2021deep}. The phase diagrams of block copolymers can be predicted by combining deep learning technique and SCFT \cite{nakamura2020phase,takeshi2021deep}. 
	
	In this work, we propose a new neural network to address inverse design problem based on the idea of multi-categorization. 
	We take AB diblock copolymer system as an example to demonstrate the performance of our network. The training and test data sets are generated from the Landau-Brazovskii (LB) model \cite{brazovskiui1996phase}.
	LB model is an effective tool to describe the phases and phase transition of diblock copolymers \cite{leibler1980theory, fredrickson1987fluctuation, shi1996theory, miao2008fluctuation, mcclenagan2019landau}.
	Let $\phi(\bm{r})$ be the order parameter, a function of spatial position $\bm{r}$, which represents the density distribution of diblock copolymers.
	The free energy functional of LB model is
		\begin{equation}
			E(\phi(\bm{r}))=\frac{1}{|\Omega|} \int_{\Omega}\left\{\frac{\xi^2}{2}[(\Delta+1) \phi]^2+\frac{\tau}{2 !} \phi^2-\frac{\gamma}{3 !} \phi^3+\frac{1}{4 !} \phi^4\right\} d \bm{r},
			\label{eq:lb}
		\end{equation}
	$\phi(\bm{r})$ satisfies the mass conservation  $\frac{1}{|\Omega|}\int_{\Omega}\phi(\bm{r}) \mathrm{d} \bm{r}=0$. $\Omega$ is the system volume. 
	The model parameters in (\ref{eq:lb}) are associated to physical conditions of diblock coplymers. Concretely, 
	$\tau$ is a temperature-like parameter related to the Flory-Huggins interaction parameter, the degree of polymerization $N$, and the A monomer fraction $f$ of each diblock copolymer chain. 
	$\tau$ can control the onset of the order-disorder spinodal decomposition. The disordered phase becomes unstable at $\tau = 0$. $\gamma$ is associated with $f$ and $N$, it's nonzero only if AB diblock copolymers chain is asymmetric. $\xi$ is the bare correlation length. Further relationship can be found in \cite{leibler1980theory, fredrickson1987fluctuation, shi1996theory, miao2008fluctuation, mcclenagan2019landau}.
	The stationary states of LB free energy functional correspond to ordered structures.
	
	The rest of the paper is organized as follows. In Section \ref{sec:direct}, we solve the LB model $\eqref{eq:lb}$ to obtain data set. In Section \ref{sec:network}, we present the multi-category inverse design neural network, and reciprocal-space data augmentation (RSDA) method for periodic structures. In Section \ref{sec:applicaton}, we take the diblock copolymer system confined in two dimension as an example to test the performance of our proposed inverse design neural network model. 
	In Section \ref{sec:conclusion}, we draw a brief summary of this work.
	
\section{Direct problem}\label{sec:direct}
	
Solving direct problem is optimizing LB free energy functional (\ref{eq:lb}) to obtain stationary states corresponding to ordered structures
		\begin{equation}
			\min_{\phi(\bm{r})} E(\phi(\bm{r})),\\\quad \text{s.t.}\quad \frac{1}{|\Omega|}\int_{\Omega}\phi(\bm{r}) \mathrm{d} \bm{r}=0.
			\label{eq:min}
		\end{equation}
Here we only consider periodic structures. Therefore, we can apply Fourier pseudospectral method to discretize the above optimization problem.
	
	\subsection{Fourier pseudospectral method} 

	For a periodic order parameter $\phi(\bm{r})$, $\bm{r} \in \Omega := \mathbb{T}^d = \mathbb{R}^d /\bm{A} \mathbb{Z}^d$, where $\bm{A}=\left(\bm{a}_1, \bm{a}_2, \ldots, \bm{a}_d\right) \in \mathbb{R}^{d \times d}$ is the primitive Bravis lattice.
	The primitive reciprocal lattice $\bm{B} = \left(\bm{b}_1, \bm{b}_2, \ldots, \bm{b}_d\right) \in \mathbb{R}^{d \times d}$, satisfying the dual relationship
		\begin{eqnarray}
			\bm{A}\bm{B}^{T}=2 \pi \bm{I}.
		\end{eqnarray}
	The order parameter $\phi(\bm{r})$ can be expanded as 
		\begin{eqnarray}
			\phi(\bm{r})= \sum_{\bm{k}\in \mathbb{Z}^d} \hat{\phi}\left(\bm{k}\right) e^{i (\bm{B} \bm{k})^T \bm{r}}, \quad \bm{r} \in \mathbb{T}^d,
		\end{eqnarray}
		\label{eq:FourierSeries}
	where the Fourier coefficient
		\begin{eqnarray}
			\hat{\phi}(\bm{k})=\frac{1}{|\mathbb{T}^d|} \int_{\mathbb{T}^d} \phi\left(\bm{r}\right) e^{-i (\bm{B} \bm{k})^T \bm{r}} d \bm{r},
		\end{eqnarray}
	$|\mathbb{T}^d|$ is the volume of $\mathbb{T}^d$.

	We define the discrete grid set as
		\begin{eqnarray}
			\mathbb{T}_N^d=\Big\{(r_{1,j_1},r_{2,j_2},...,r_{d,j_d})=\bm{A}\big(  j_1/N , j_2/N, ...,j_d/N \big)^T, 0 \leq j_i<N, j_i \in \mathbb{Z}, i= 1,2,...,d\Big\},\label{eq:12}
		\end{eqnarray}
	where the number of elements of $\mathbb{T}_N^d$ is $M = N^d$. 
	Denote grid periodic function space $\mathcal{G}_N$ = \big\{$f:\mathbb{T}^d_N \mapsto \mathbb{C}$, $f$ is periodic \big\}. For any periodic grid functions $F,G \in \mathcal{G}_N$, the $\ell^2$-inner product is defined as
		\begin{eqnarray}
			\langle F, G \rangle_N = \frac{1}{M}\sum_{\bm{r}_j \in \mathbb{T}^d_N} F(\bm{r}_j)\bar{G}(\bm{r}_j). 
		\end{eqnarray}
	The discrete reciprocal space is
		\begin{eqnarray}
			\mathbb{K}_N^d=\left\{\bm{k}=\left(k_j\right)_{j=1}^d \in \mathbb{Z}^d:-N/2 \leq k_j<N/2\right\},
		\end{eqnarray}
	the discrete Fourier coefficients of $\phi(\bm{r})$ in $\mathbb{T}_N^d$ can be represented as
		\begin{eqnarray}
			\hat{\phi}(\bm{k})= \langle \phi(\bm{r}_j), e^{i(\bm{Bk})^T\bm{r}_j}\rangle_N=\frac{1}{M} \sum_{\bm{r}_j\in \mathbb{T}_N^d} \phi\left(\bm{r}_j\right) e^{-i (\bm{B} \bm{k})^T \bm{r}_j}, \quad \bm{k}\in \mathbb{K}_N^d.
		\end{eqnarray}
	For $\bm{k} \in \mathbb{Z}^d$, and $\bm{l} \in \mathbb{Z}^d$, we have the discretize orthogonality,
		\begin{eqnarray}
			\langle e^{i (\bm{Bk})^T \bm{r_j}},e^{i (\bm{Bl})^T \bm{r_j}} \rangle_N = \left\{
			\begin{array}{cc}
				1, & \bm{k} = \bm{l}+N\bm{m}, ~\bm{m} \in \mathbb{Z}^d, \\
				0, & otherwise. 
			\end{array} \right.
			\label{orthonormal}
		\end{eqnarray} 
	Therefore, the discrete Fourier transform of $\phi(\bm{r_j})$ is,
		\begin{eqnarray}
			\phi(\bm{r}_j)= \sum_{\bm{k}\in \mathbb{K}_N^d} \hat{\phi}\left(\bm{k}\right) e^{i (\bm{B} \bm{k})^T \bm{r}_j}, \quad \bm{r}_j \in \mathbb{T}_N^d.
		\end{eqnarray}
 The $N^d$-order trigonometric polynomial is
		\begin{eqnarray}
			I_N \phi(\bm{r})=
			\sum_{\bm{k}\in \mathbb{K}_N^d} \hat{\phi}\left(\bm{k}\right) e^{i (\bm{B} \bm{k})^T \bm{r}},\quad \bm{r}\in \mathbb{T}^d.
			\label{eq:triPoly}
		\end{eqnarray}
	Then for $\bm{r}_j\in \mathbb{T}_N^d $ , we have $\phi(\bm{r}_j) \approx I_N \phi(\bm{r}_j)$.
	
	Due to the orthogonality $\eqref{orthonormal}$, the LB energy functional $E(\phi)$ can be discretized as 
		\begin{eqnarray}
			\begin{split}
				E_h[\hat{\Phi}] &=\frac{\xi^2}{2} \sum_{\bm{h}_1+\bm{h}_2=\bm{0}}\left[1-\left(\bm{B} \bm{h}_1\right)^T\left(\bm{B} \bm{h}_2\right)\right]^2 \hat{\phi}\left(\bm{h}_1\right) \hat{\phi}\left(\bm{h}_2\right)+\frac{\tau}{2 !} \sum_{\bm{h}_1+\bm{h}_2=\bm{0}} \hat{\phi}\left(\bm{h}_1\right) \hat{\phi}\left(\bm{h}_2\right) \\
				&-\frac{\gamma}{3 !} \sum_{\bm{h}_1+\bm{h}_2+\bm{h}_3=\bm{0}} \hat{\phi}\left(\bm{h}_1\right) \hat{\phi}\left(\bm{h}_2\right) \hat{\phi}\left(\bm{h}_3\right)\\
				&+\frac{1}{4 !} \sum_{\bm{h}_1+\bm{h}_2+\bm{h}_3+\bm{h}_4=\bm{0}} \hat{\phi}\left(\bm{h}_1\right) \hat{\phi}\left(\bm{h}_2\right) \hat{\phi}\left(\bm{h}_3\right) \hat{\phi}\left(\bm{h}_4\right),
			\end{split}
			\label{eq:E}
		\end{eqnarray}
	where $\bm{h}_i \in \mathbb{K}_N^d$, $ i=1,2,3,4$, and $\hat{\Phi}=\left(\hat{\phi}_1, \hat{\phi}_2, \ldots, \hat{\phi}_M\right)^T \in \mathbb{C}^M$. The convolutions in the above expression can be efficiently calculated through the fast Fourier transform (FFT). Moreover, the mass conservation constraint $\frac{1}{|\Omega|}\int_{\Omega}\phi(\bm{r}) \mathrm{d} \bm{r}=0$ is discretized as
		\begin{eqnarray}
			e^{T}\hat{\Phi} = 0,
		\end{eqnarray}
	where $e = (1,0,...,0)^{T}\in \mathbb{R}^{M}$. Therefore, (\ref{eq:min}) reduces to a finite dimensional optimization problem
		\begin{eqnarray}
			\min _{\hat{\Phi} \in \mathbb{C}^{\mathrm{M}}} E_h[\hat{\Phi}]=
			G_h[\hat{\Phi}]+F_h[\hat{\Phi}], \quad s.t.\quad e^{T}\hat{\Phi} = 0,
			\label{eq:Min}
		\end{eqnarray}
	where $G_h$ and $F_h$ are discretize interaction and bulk energies
		\begin{eqnarray}
			\begin{split}
				G_{\bm{h}}(\hat{\Phi})
				&= \frac{\xi^2}{2} \sum_{\bm{h}_1+\bm{h}_2=\bm{0}}\left[1-\left(\bm{B} \bm{h}_1\right)^T\left(\bm{B} \bm{h}_2\right)\right]^2 \hat{\phi}\left(\bm{h}_1\right) \hat{\phi}\left(\bm{h}_2\right), \\
				F_{\bm{h}}(\hat{\Phi})
				&= \frac{\tau}{2 !} \sum_{\bm{h}_1+\bm{h}_2=\bm{0}} \hat{\phi}\left(\bm{h}_1\right) \hat{\phi}\left(\bm{h}_2\right) -\frac{\gamma}{3 !} \sum_{\bm{h}_1+\bm{h}_2+\bm{h}_3=\bm{0}} \hat{\phi}\left(\bm{h}_1\right) \hat{\phi}\left(\bm{h}_2\right) \hat{\phi}\left(\bm{h}_3\right)\\
				&+\frac{1}{4 !} \sum_{\bm{h}_1+\bm{h}_2+\bm{h}_3+\bm{h}_4=\bm{0}} \hat{\phi}\left(\bm{h}_1\right) \hat{\phi}\left(\bm{h}_2\right) \hat{\phi}\left(\bm{h}_3\right) \hat{\phi}\left(\bm{h}_4\right).
			\end{split}
			\label{eq:gf}
		\end{eqnarray}
	In the work, we employ the adaptive APG method to solve \eqref{eq:Min}.
	
	\subsection{Phase diagram}
	
\begin{figure}[!htbp]
		\centering
		\includegraphics[width=13.5cm]{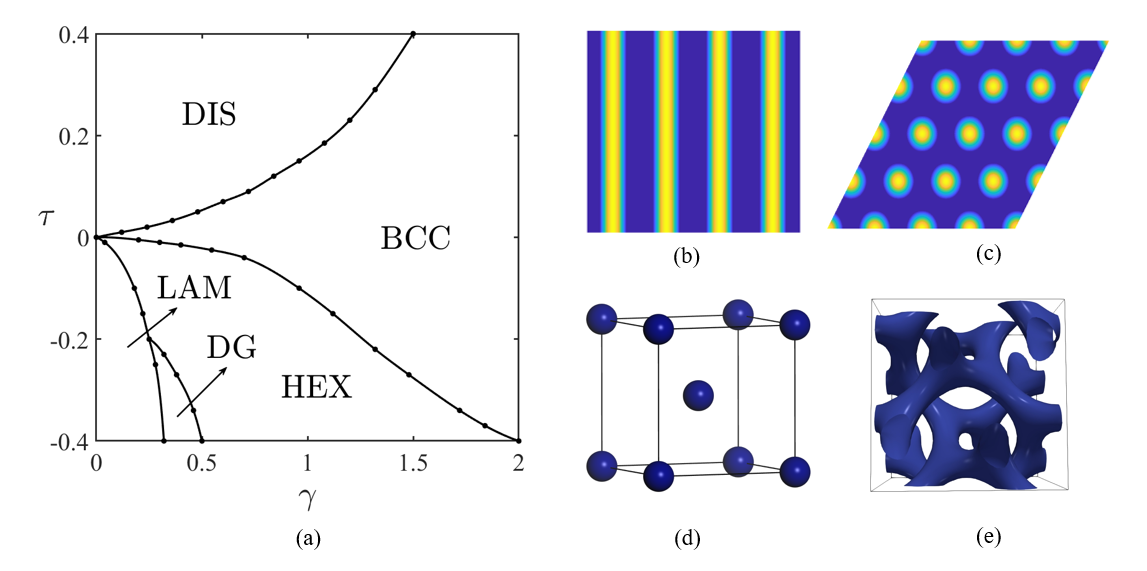}
		\caption{(a) Phase diagram of Landau-Brazovskii model. (b) Lamellar phase (LAM). (c) Cylindrical hexagonal phase (HEX). (d) Body-centered cubic phase. (e) Double gyroid phase.\label{fig:P}}
	\end{figure}

	Given parameters $[\gamma, \tau]$, we can obtain the stationary states by solving the free energy functional $\eqref{eq:gf}$. 
	Due to the non-convexity of LB free energy functional, there are many, even infinite stationary states for given parameters. We need to determine the stationary state with the lowest energy which corresponds to the most probable ordered structure observed in experiments. It requires to comparing the energies of stationary states to obtain the stable structure and constructing phase diagram. In the following, we consider disordered, cylindrical hexagonal (HEX), lamellar (LAM), body-centered cubic, and double gyroid phases as candidate structures. We use the AGPD software \cite{jiang2022agpd} to produce a $(\tau, \gamma)$-plane phase diagram, as shown in Figure~\ref{fig:P}.  
	The obtained phase diagram is consistent with previous work \cite{shi1999nature,zhang2008an,mcclenagan2019landau}.
	
\section{Inverse design neural network}
\label{sec:network}

\subsection{Multi-category inverse design network}
\begin{figure}[!htbp]
\centering
\includegraphics[width = 14.5cm]{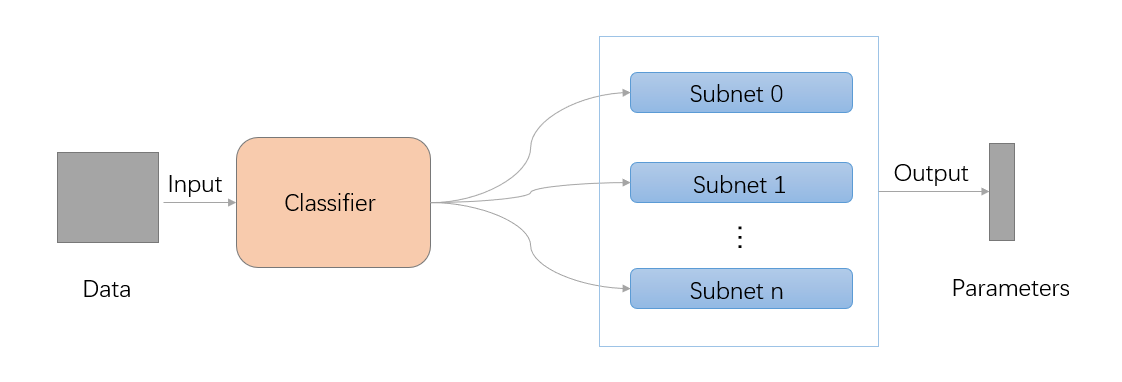}
\caption{Multi-category inverse design network.}
\label{fig:FC}
\end{figure}

The architecture of multi-category inverse design network for predicting the physical parameters for desired periodic structure, is shown in Figure~\ref{fig:FC}. The neural network mainly consists of two modules: a classifier and SPM networks. The former (orange block) is identifying and classifying candidate structures, and the latter (blue block) including a family of subnets, is a mapping connecting physical parameters and ordered structures. 

\begin{figure}[!htbp]
\centering
\includegraphics[width = 13.5cm]{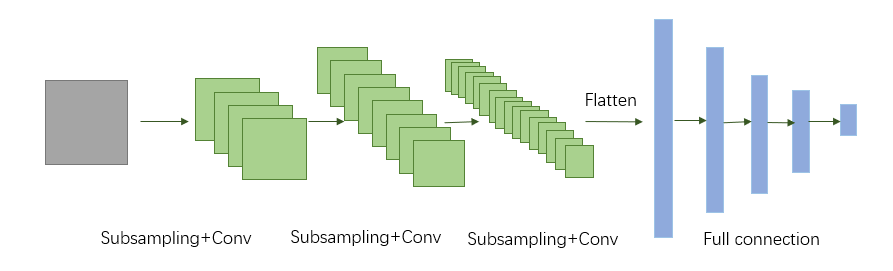}
\caption{Network architecture of classifier and each SPM subnet.}
\label{fig:CF}
\end{figure}

According to the characteristics of problem, we can design corresponding network architectures of classifier and SPM subnets. In this work, the classifier and each SPM subnet use the same network architectures as shown in Figure~\ref{fig:CF}. Concretely, the network contains an input layer, three convolutional layers, three MaxPooling layers, four fully connected layers, and an output layer. For the classifier network, the size of the output layer represents the number of categories, while for each subnet, it means the number of predicted physical parameters.
The subnet architecture is a development of Lenet-5 network \cite{lecun1998gradient}. 

\subsection{Reciprocal-space data augmentation (RSDA) method}
Generally, the ability of a network depends not only on its architecture, but also on the amount and properties of data. 
Due to the rotation and translation invariance of periodic structures, how to make the classifier network recognize the invariance is very important. 
Here we use data augmentation method to increase the amount of data. Existing data augmentation methods, such as Swapping, Mixup, Insertion, Substitution, are often used for classification tasks, see a recent review \cite{li2021data} and reference therein. However, as data set increases, these data augmentation approaches might become expensive, and be easy to use wrong labels to train the network. 
For one-dimensional periodic phases, Yao et al. \cite{yao2021deep} used Period-Filling method for data augmentation. However, this approach is difficult to extend to higher dimensions.

In this paper, we propose an extensible data augmentation method implemented in reciprocal space, called RSDA method. 
Denote that the fundamental domain of periodic structure $\phi(\bm{r})$ is $\Omega=\{\sum_{j=1}^d \zeta_j \bm{a}_j,  \bm{a}_j\in\mathbb{R}^d,  \zeta_j \in[0,1)\}$. For any translation $\bm{t}\in \mathbb{R}^d$, $\bm{t}/\bm{A}\mathbb{Z}^d \in \Omega$, therefore, we only need to consider the translation in fundamental domain $\Omega$. From the Fourier expansion $\eqref{eq:triPoly}$, and for rotation matrix $\bm{R} \in SO(d)$, $\bm{t} \in \Omega$, we have 
\begin{equation}
    \begin{aligned}
        \phi(\bm{R} \bm{r}+\bm{t})&=\sum_{ \bm{h} \in \mathbb{K}_N^d} \hat{\phi}(\bm{h}) e^{i(\bm{B} \bm{h})^T(\bm{R} \bm{r}+\bm{t})}
        \\
		&=\sum_{ \bm{h} \in  \mathbb{K}_N^d} \hat{\phi}(\bm{h})e^{i\left(\bm{B} \bm{h}\right)^T \bm{t}} e^{i\left(\tilde{\bm{B}} \bm{h}\right)^T \bm{r}} 
		\\
		&=\sum_{ \bm{h} \in  \mathbb{K}_N^d} \tilde{ \phi}(\bm{h}) e^{i\left(\tilde{\bm{B}} \bm{h}\right)^T \bm{r}} 
    \end{aligned}
\end{equation}
where $\tilde{\bm{B}} = \bm{R}^T \bm{B}$ is a new reciprocal lattice by rotation transformation, and $\tilde{\phi}(\bm{h}) =\hat{\phi}(\bm{h})e^{i\left(\bm{B} \bm{h}\right)^T \bm{t}}$ is a new Fourier coefficient associated with translation transformation. 
Obviously, the RSDA method is easy to implement, and can be suitable for arbitrary dimensional periodic structures.

\section{Application}
\label{sec:applicaton}

\begin{figure}[!htbp]
\centering
\includegraphics[width = 13cm]{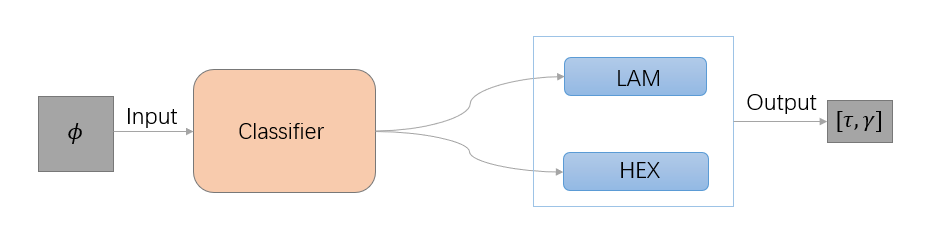}
\caption{The network of two-dimensional diblock copolymer system.}
\label{fig:FC0}
\end{figure}

In this section, we apply the multi-category network model to diblock copolymers confined on two-dimensional plane to obtain the mapping from periodic structures to physical parameters. As phase diagram shows, in two dimension, only LAM and HEX phases are thermodynamic stable. Therefore, the SPM has two subnets, i.e., LAM and HEX networks. As shown in Figure~\ref{fig:FC0}, the classifier is used to distinguish LAM or HEX phase after inputting the order parameter $\phi$. For desired structures, LAM and HEX subnets is used to predict the physical parameters $(\tau^{*}, \gamma^{*})$. Dataset required for the classifier consists of $\{(\phi_{j}, y_{j})\}_{j=1}^{N}$, $y_{j}$ is the label, and for each SPM subnet consists of $\{(\phi_{j}, (\tau^{*}_j, \gamma^{*}_j))\}_{j=1}^{N}$,  $N$ is the size of training data. 

Table~\ref{Tab:2} shows these network parameters of classifier and SPM subnets. In all networks, the size of the output layer is 2. We adopt the ReLU function as the activation function and the Adam optimizer with a learning rate of $10^{-4}$ to train neural network model. Kaiming\_uniform \cite{he2015delving} is used to initialize the network parameters both in classifier and subnets. We set the maximum epoches to be 20000 and stop training if the error on the validation set decreases to $10^{-5}$ to prevent overfitting. For the classifer, the loss function is defined as 

\begin{equation}
    \begin{aligned}
L=\frac{1}{N} \sum_{i=1}^{N}-\left[y_i \cdot \log \left(p_i\right)+\left(1-y_i\right) \cdot \log \left(1-p_i\right)\right],
    \end{aligned}
\end{equation}
where $y_i$ represents the label of the sample, LAM phase is 0, and HEX phase is 1. $p_i$ is the probability of identifying HEX phase.
For each SPM subnet, the loss function is 
\begin{equation}
    \begin{aligned}
M S E=\frac{1}{N} \sum_{i=1}^N\left(\bm{u}_i-\tilde{\bm{u}}_i\right)^2,
    \end{aligned}
\end{equation}
where $\bm{u} = (\tau^{*},\gamma^{*})$ is targeted parameters and $\tilde{\bm{u}} = (\tau,\gamma)$ is predicted parameters. 

\begin{figure}[!htbp]
  \centering
  \subfigure[$\Omega_{LAM}$]{
  \centering
    \includegraphics[width=2.5cm]{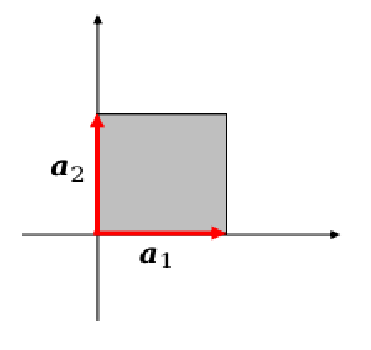}
  }
  \quad
  \subfigure[$\Omega_{HEX}$]{
  \centering
    \includegraphics[width=2.5cm]{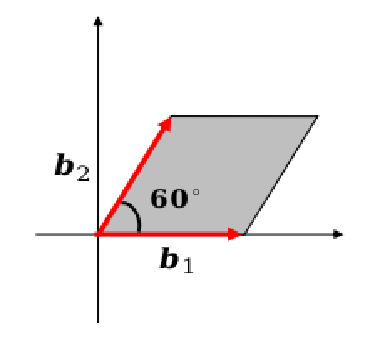}
  }
  \quad
  \subfigure[\scriptsize{Rotation of LAM}]{
  \centering
    \includegraphics[width=2.5cm]{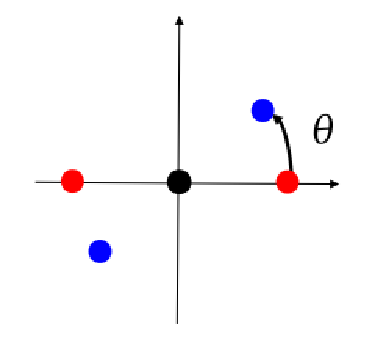}
  }
  \quad
 \subfigure[\scriptsize{Rotation of HEX}]{
  \centering
    \includegraphics[width=2.5cm]{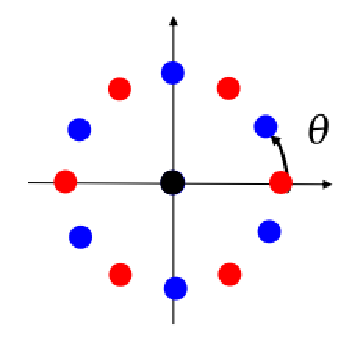}
  }
  \caption{Schematic diagram of fundamental domain (a) LAM phase $\Omega_{LAM}$; (b) HEX phase $\Omega_{HEX}$. Illustration of rotating primary spectral points of (c) LAM, and (d) HEX phases in reciprocal space. Reference state (red), and transform state after rotating $\theta$ degree (blue).}
  \label{Fig:basedvector}
\end{figure}

\begin{table}[!htbp] 
\caption{Network parameters of classifier and each subnet. The notations of parameters in the table are: in channels (i), out channels (o), kernel size (k), stride (s), padding (p), batch size (Nb).\label{Tab:2}}
\newcolumntype{C}{>{\centering\arraybackslash}X}
\begin{tabularx}{\textwidth}{p{6.5cm}CCC}
\toprule
\textbf{Layer type}	& \textbf{Output shape}	& \textbf{Layer type}     & \textbf{Output shape}\\
\midrule
Conv2d $(\mathrm{i}=1, \mathrm{o}=4, \mathrm{k}=5, \mathrm{p}=2, \mathrm{~s}=1)$& $\left(\mathrm{N}_{\mathrm{b}}, 4,40,40\right)$ & & Connect left \\
MaxPool2d $(\mathrm{k}=2, \mathrm{~s}=2)$ & $\left(\mathrm{N}_{\mathrm{b}}, 4,20,20\right)$ & Reshape & $\left(\mathrm{N}_{\mathrm{b}}, 256\right)$ \\
Conv2d $(\mathrm{i}=4, \mathrm{o}=8, \mathrm{k}=5, \mathrm{p}=2, \mathrm{~s}=1)$ & $\left(\mathrm{N}_{\mathrm{b}}, 8,20,20\right)$ & Fully connected & $\left(\mathrm{N}_{\mathrm{b}}, 128\right)$ \\
MaxPool2d $(\mathrm{k}=2, \mathrm{~s}=2)$ & $\left(\mathrm{N}_{\mathrm{b}}, 8,10,10\right)$ & Fully connected & $\left(\mathrm{N}_{\mathrm{b}}, 64\right)$ \\
Conv2d $(\mathrm{i}=8, \mathrm{o}=16, \mathrm{k}=5, \mathrm{p}=1, \mathrm{~s}=1)$ & $\left(\mathrm{N}_{\mathrm{b}}, 16,8,8\right)$ & Fully connected & $\left(\mathrm{N}_{\mathrm{b}}, 10\right)$ \\
MaxPool2d $(\mathrm{k}=5, \mathrm{~s}=1)$ & $\left(\mathrm{N}_{\mathrm{b}}, 16,4,4\right)$ & Fully connected & $\left(\mathrm{N}_{\mathrm{b}}, 2\right)$
\\
\bottomrule
\end{tabularx}
\end{table}
It is well known that LAM and HEX phases have 2- and 6-fold symmetries, respectively.  As shown in  Figure~\ref{Fig:basedvector}(a-b), the fundamental domains  $\Omega_{LAM}$ of LAM phase is a square, while the fundamental domain $\Omega_{HEX}$ of HEX is a parallelogram region due to the 6-fold rotational symmetry. Their corresponding primitive Bravis lattice is  
$$
\bm{A}_{LAM} =\lambda_1 \left[\begin{array}{cc}
1 &0  \\
0 & 1  \\
\end{array}\right],\quad \bm{A}_{HEX} =\lambda_2 \left[\begin{array}{cc}
1 & \cos(\pi/3)  \\
0 & \sin(\pi/3)  \\
\end{array}\right].
$$
Lattice constant $\lambda_1, \lambda_2$ depends on the model parameters.

Now we construct dataset of LAM phase. We discrete the $\tau$-$\gamma$ phase space of stable LAM phase with step size $[\Delta \tau, \Delta \gamma]= [0.008, 0.02]$ to form dataset $S^1_{LAM}$ for training and validating, and with step size $[\Delta \tau, \Delta \gamma]= [0.003, 0.015]$ to form $S^2_{LAM}$ for testing. We randomly choose 627 parameter pairs as a group $G^1_{LAM}$ from $S^1_{LAM}$ to construct training and validation sets, and 262 parameter pairs as a group $G^2_{LAM}$ from $S^2_{LAM}$ to build the test set. 
Then we generate LAM phases by solving direct problem with each selected parameter pair in $G^2_{LAM}$ and $G^2_{LAM}$. 

We augment data by rotation and translation. The rotation matrix is 
\begin{align}
\bm{R} = \left[\begin{array}{cc}
\cos\theta &\sin \theta  \\
-\sin \theta & \cos \theta \\
\end{array}\right], \label{eq:rotationmatrix}
\end{align}
where $\theta$ is the rotation angle.
LAM phases have 2-fold rotational symmetry, therefore, $\theta\in [0,\pi)$.
Figure \ref{Fig:basedvector}\,(c) gives a sketch plot of rotating LAM phase in the reciprocal space. Concretely, we rotate 627 LAM phases in $G^1_{LAM}$ with $\theta = \{j\pi/60\}_{j=1}^{60}$ and translate them with $\bm{t} = (i,i)^{\top}\in\Omega_{LAM}, i\in \{0, 0.2, 0.4, 0.6, 0.8\}$. The input dataset we have obtained includes 188100 LAM phases. We split randomly these samples into a training set and a validation set in a ratio of 4:1.
In the group $G^2_{LAM}$, 262 LAM phases are rotated by $\theta \in \{3\degree$, $7\degree$, $19\degree$, $20\degree$, $27\degree$, $34\degree$, $78\degree$, $80\degree$, $82\degree$, $114\degree\}$ and translated by $\bm{t} = (i,i)^{\top}, i\in \{0.17$, $0.37$, $0.51$, $0.65$, $0.73\}$ as test set. 

Similarly, we construct a dataset of HEX phase.
The $\tau$-$\gamma$ domain of the stable HEX phase is discretized with step size $[\Delta \tau, \Delta \gamma]= [0.008, 0.02]$ to form $S^1_{HEX}$ for training and validating, and with step size $[\Delta \tau, \Delta \gamma]= [0.003, 0.015]$ to form $S^2_{HEX}$ for testing. We randomly choose $1600$ parameter pairs as a group $G^1_{HEX}$ from $S^1_{HEX}$ to construct training and validation sets, and 300 parameter pairs as a group $G^2_{HEX}$ from $S^2_{HEX}$ to build test set. 
We still obtain HEX phases by solving direct problem for selected parameters. 

Then we augment dataset by rotation and translation operators. 
The HEX has 6-fold rotational symmetry, therefore, the rotation angle $\theta$ in (\ref{eq:rotationmatrix}) belongs to $[0, \pi/3)$. An illustration of rotating HEX phase in the reciprocal space is shown in Figure~\ref{Fig:basedvector}\,(d). 
Concretely, we discretize rotation angle $\theta = \{j\pi/60\}_{j=1}^{20}$ and select translation vector $\bm{t} = (i,i)^{\top}$ in $\Omega_{HEX}$, $i\in \{0, 0.2, 0.4, 0.6, 0.8\}$, for processing each sample in $G^1_{HEX}$. It results in 16000 HEX phases. We randomly divide them into a training set and a validation set with a ratio of 4:1. The generated dataset of 300 in $G^2_{HEX}$ are rotated by $\theta \in \{3\degree$, $11\degree$, $24\degree$, $27\degree$, $38\degree$, $40\degree$, $52\degree$, $54\degree$, $59\degree\}$ and translated by $\bm{t} = (i,i)^{T}$, $i \in \{0.17$, $0.37$, $0.51$, $0.65$, $0.73\}$ as test set. 

\begin{figure}[!htbp]
  \centering
  \subfigure[]{
  \centering
    \includegraphics[width=2.5cm]{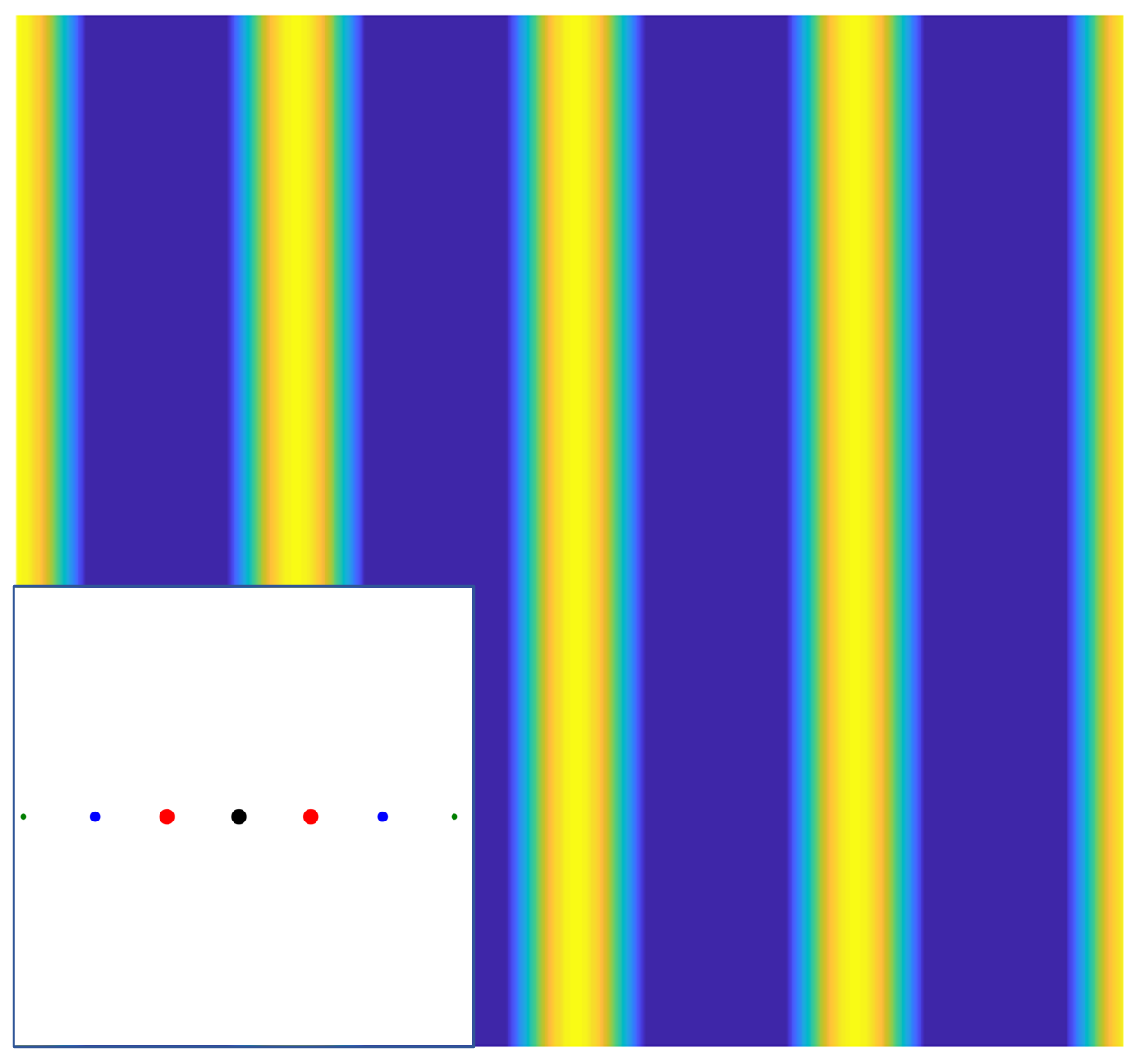}
  }
  \quad\quad
  \subfigure[]{
  \centering
    \includegraphics[width=2.4cm]{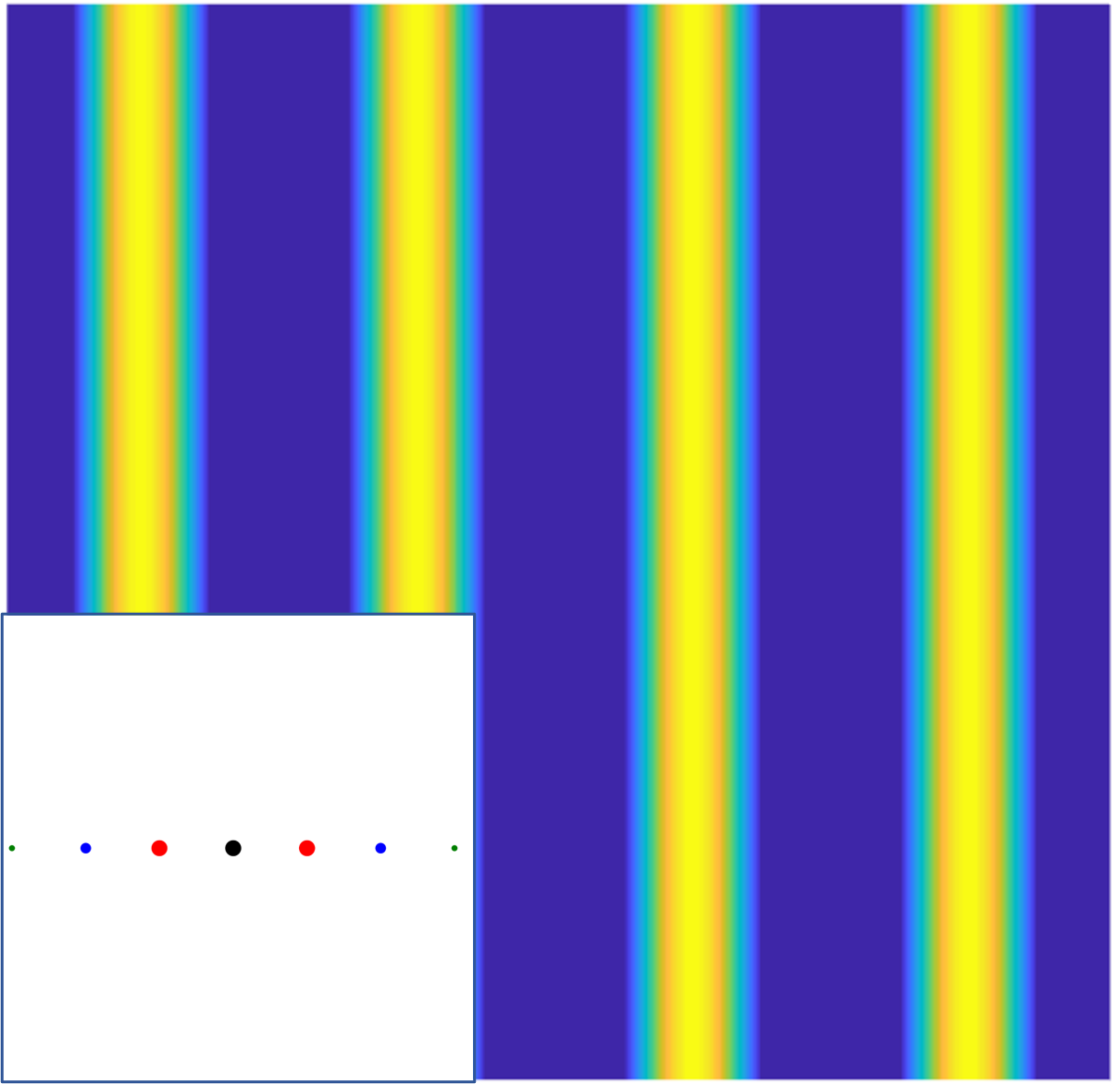}
  }
  \quad\quad
  \subfigure[]{
  \centering
    \includegraphics[width=2.4cm]{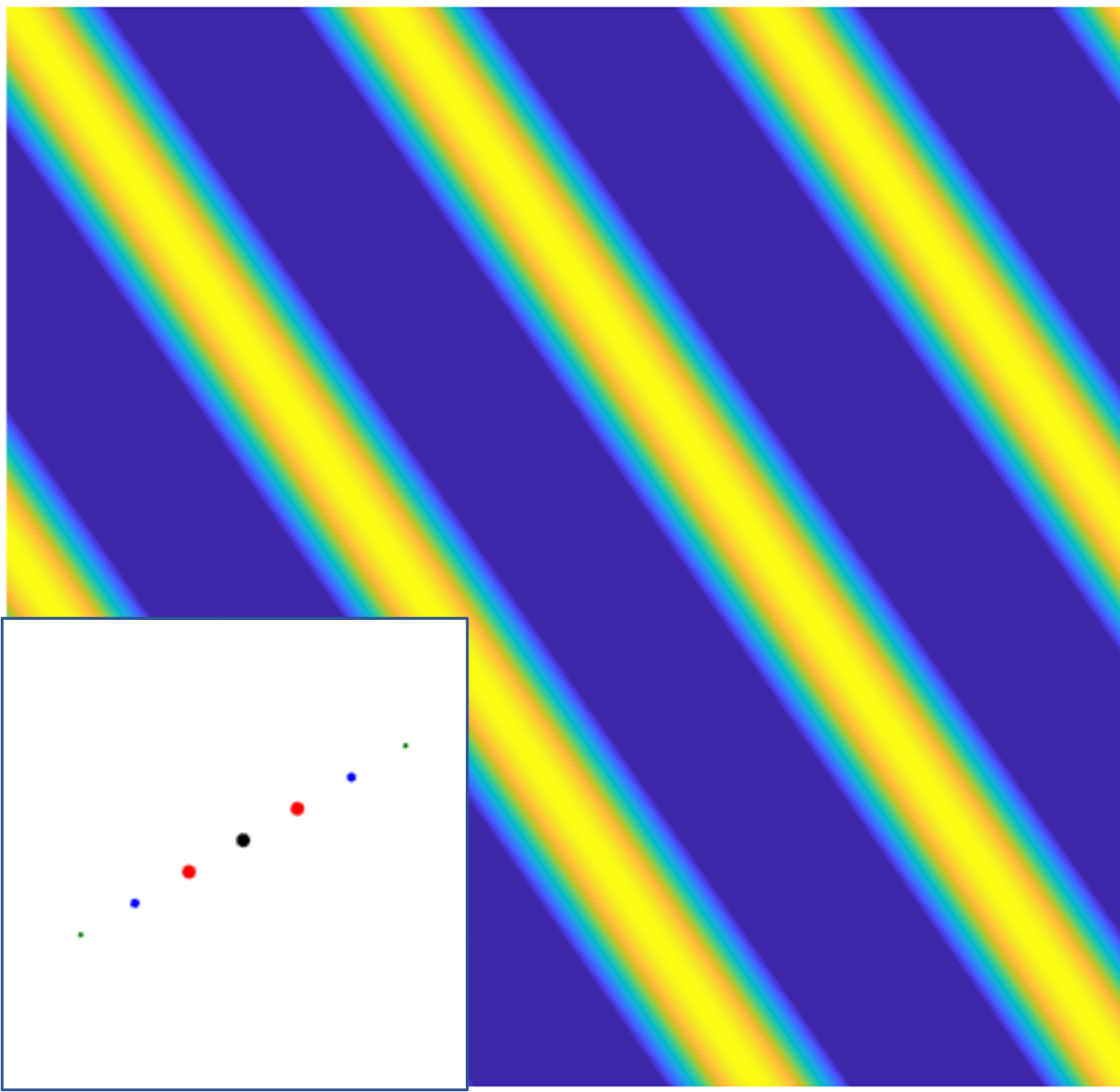}
  }
  \quad\quad
 \subfigure[]{
  \centering
    \includegraphics[width=2.4cm]{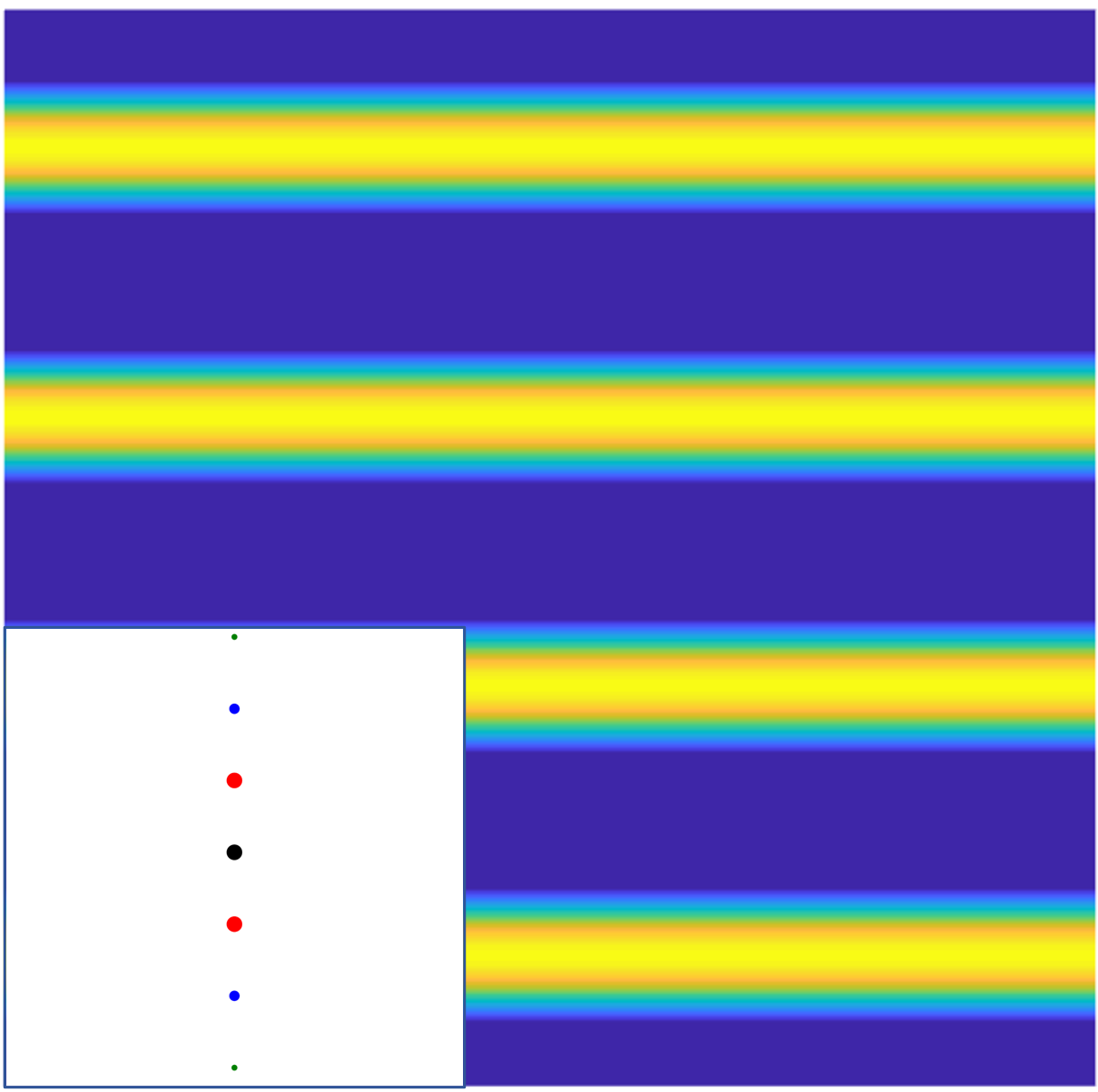}
  }\\
  \subfigure[]{
  \centering
    \includegraphics[width=3.2cm]{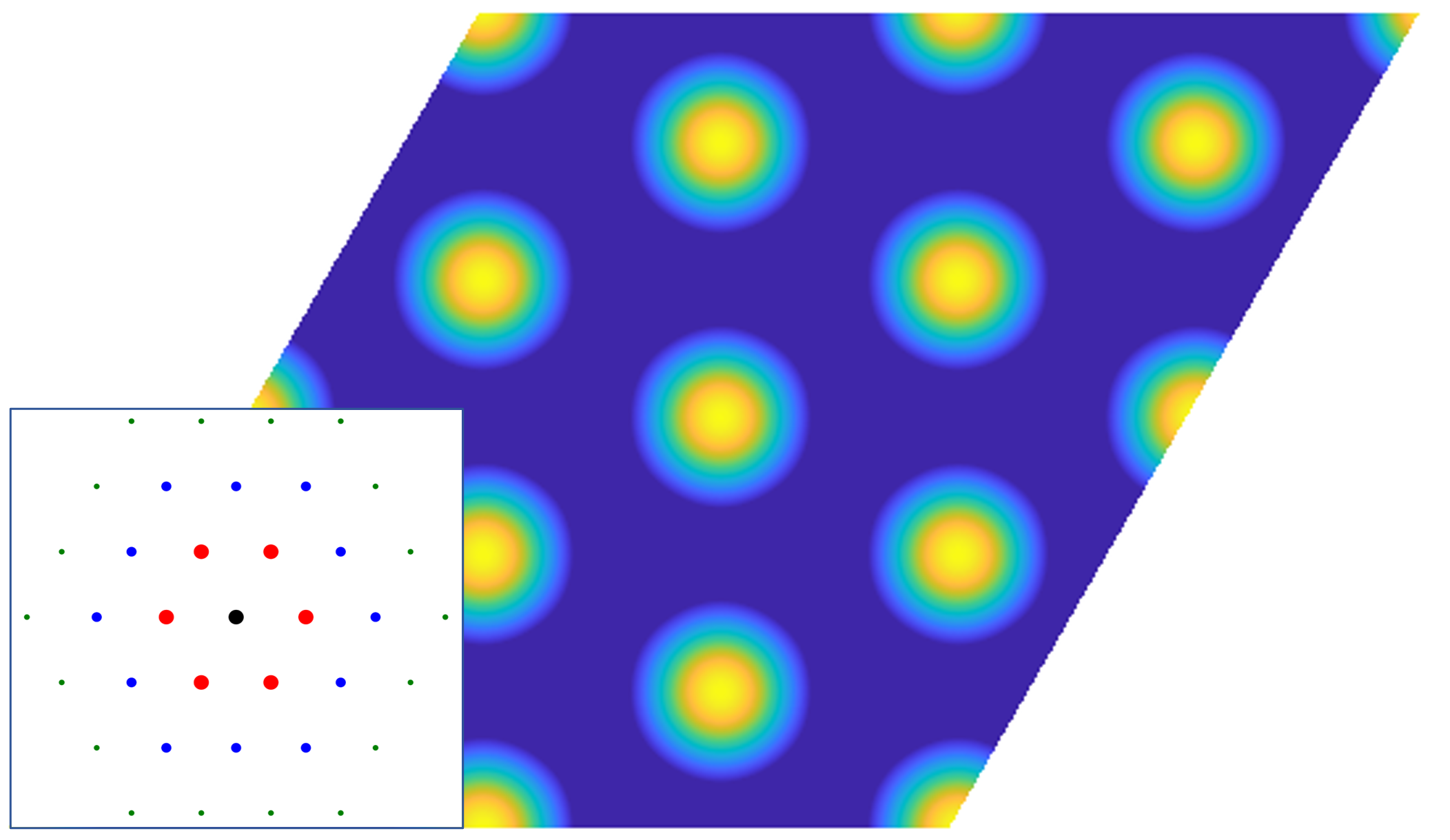}
  } 
  \subfigure[]{
  \centering
    \includegraphics[width=3.2cm]{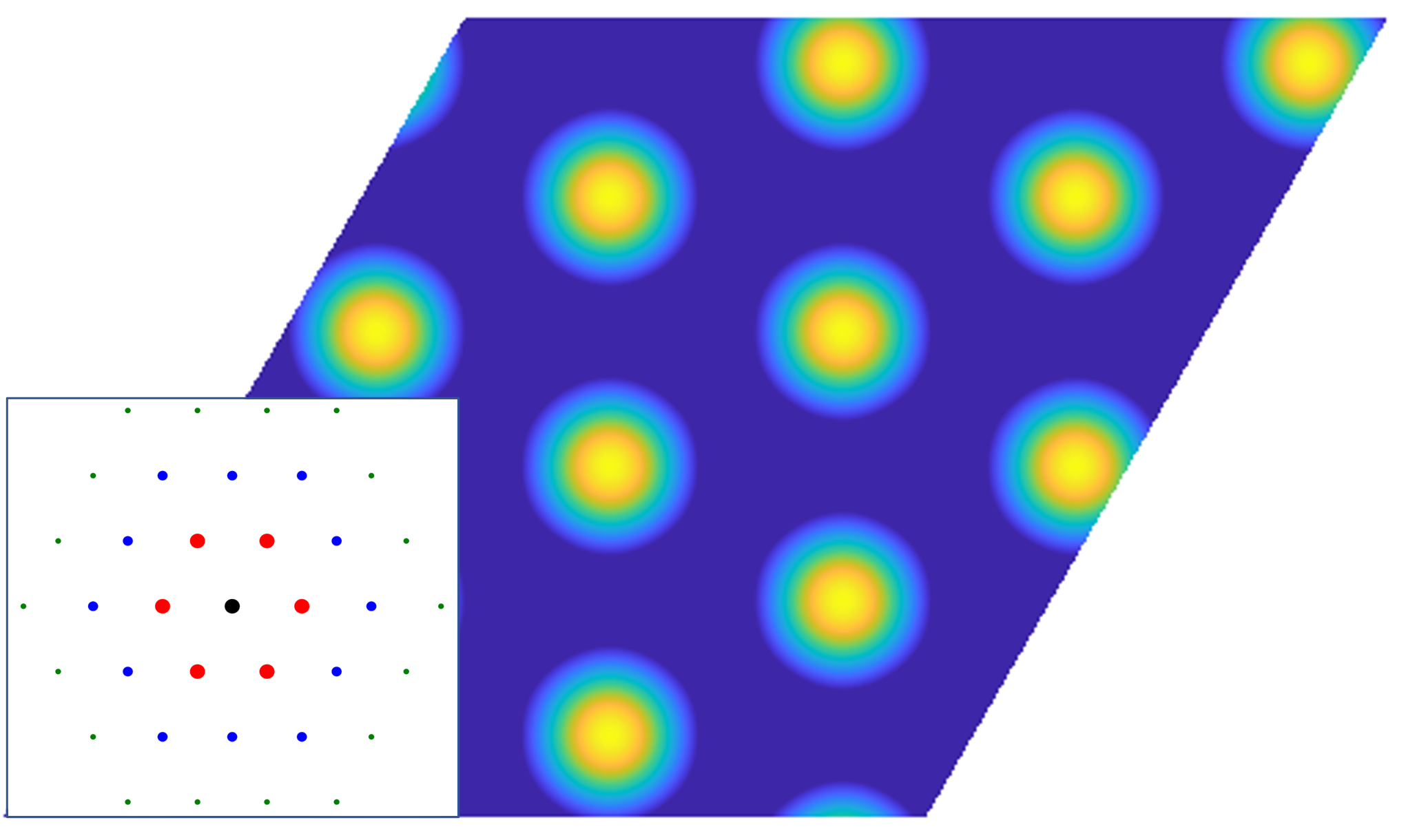}
  }  
    \subfigure[]{
  \centering
    \includegraphics[width=3.2cm]{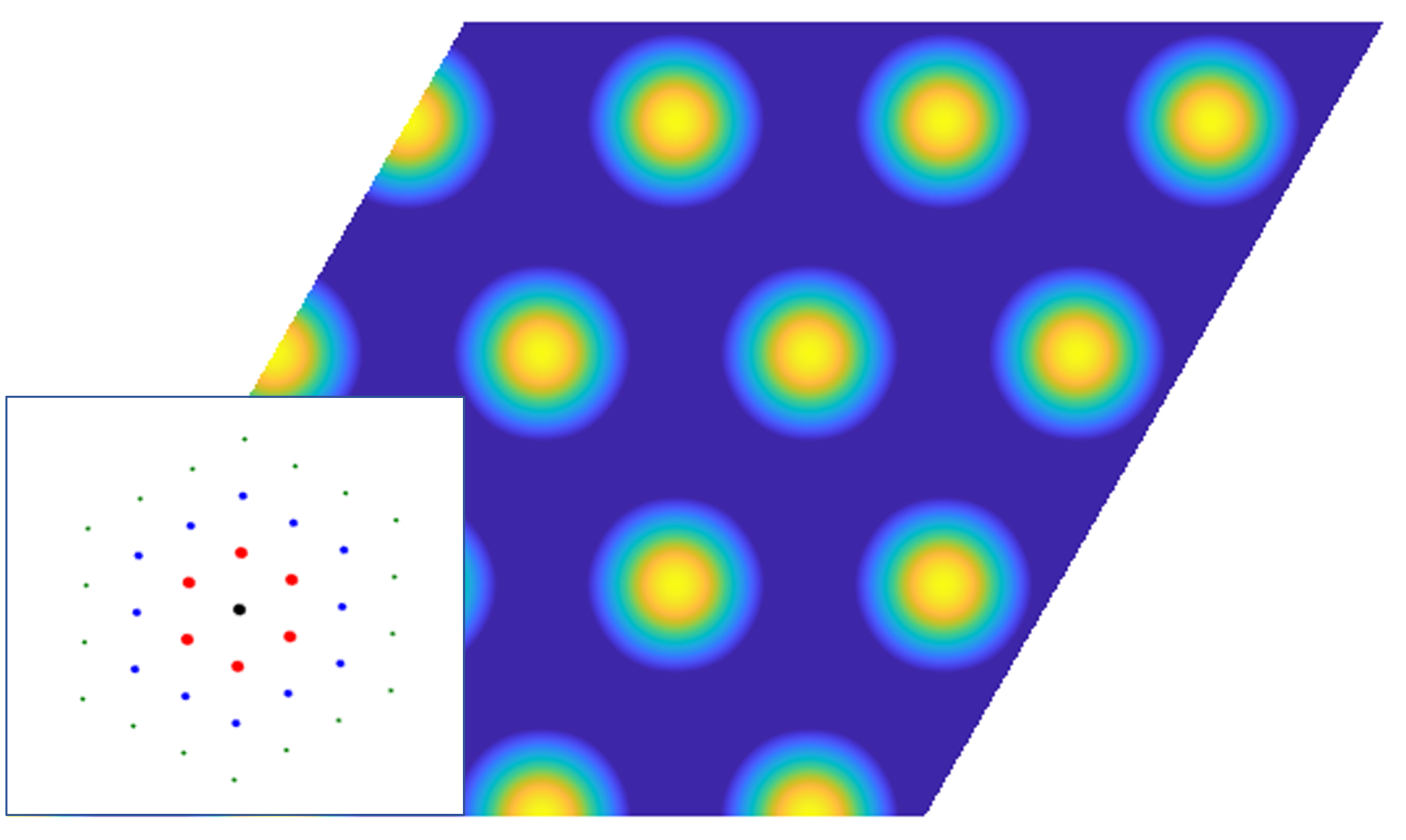}
  }
    \subfigure[]{
  \centering
    \includegraphics[width=3.1cm]{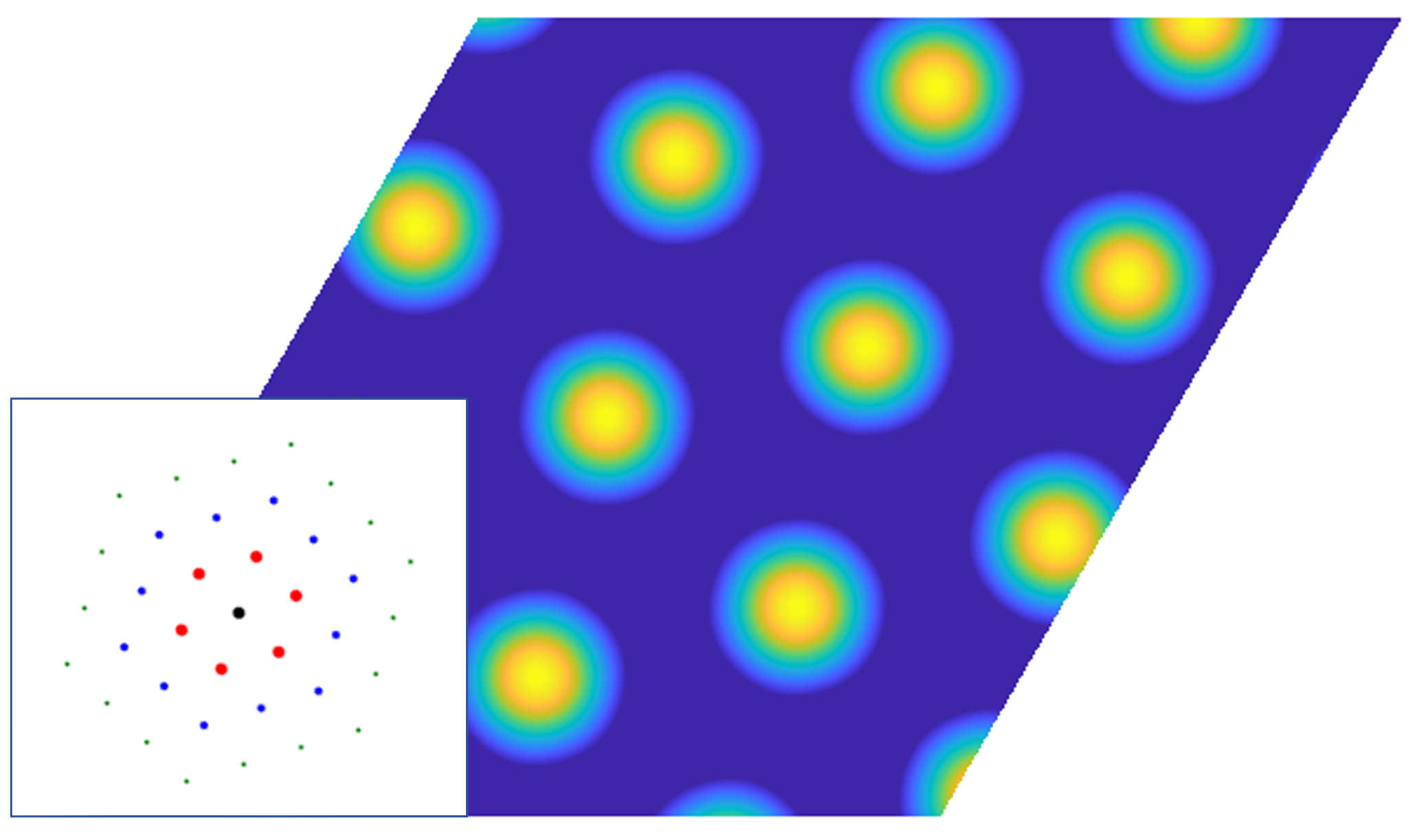}
  }
  \caption{Density and spectral pattern of LAM and HEX phases under rotation and translation transformation.
  LAM phase at $[\tau, \gamma]=[-0.3, 0.2]$: (a) $\theta$ = 0, $\bm{t}= (0,0)^{\top}$; (b) $\theta$ = 0, $\bm{t} = (0.1,0.1)^{\top}$; (c) $\theta$ = $\pi/6$, $\bm{t} = (0,0)^{\top}$; (d) $\theta$ = $\pi/2$, $\bm{t} = (0.1,0.1)^{\top}$. HEX phase $[\tau, \gamma]=[-0.3, 0.8]$: (e) $\theta$ = 0, $\bm{t} = (0,0)^{\top}$; (f) $\theta$ = 0, $\bm{t} = (0.1,0.1)^{\top}$; (g) $\theta$ = $\pi/6$, $\bm{t}= (0,0)^{\top}$; (h) $\theta$ = $\pi/4$, $\bm{t} = (0.1,0.1)^{\top}$. }
  \label{fig:lam_R}
\end{figure}

Some LAM and HEX phases under rotation and translation transformation are visually illustrated in Figure \ref{fig:lam_R}.
The size of training, validation and test sets for the classifier is given in Table \ref{Tab:split}.

\begin{table}[!htbp]
\centering
\caption{Amount of training, validation and test data for the classifier.\label{Tab:split}} 
\begin{tabular}{cccc}
\toprule[1pt] 
  & \text{Training set} & \text{Validation set} & \text{Test set} \\ 
\hline\text{Classifier}  &   278480 &  69620 &  26600\\  
\bottomrule[1pt]
\end{tabular}  
\end{table}

\begin{figure}[!htbp]
  \centering
  \includegraphics[width=11cm]{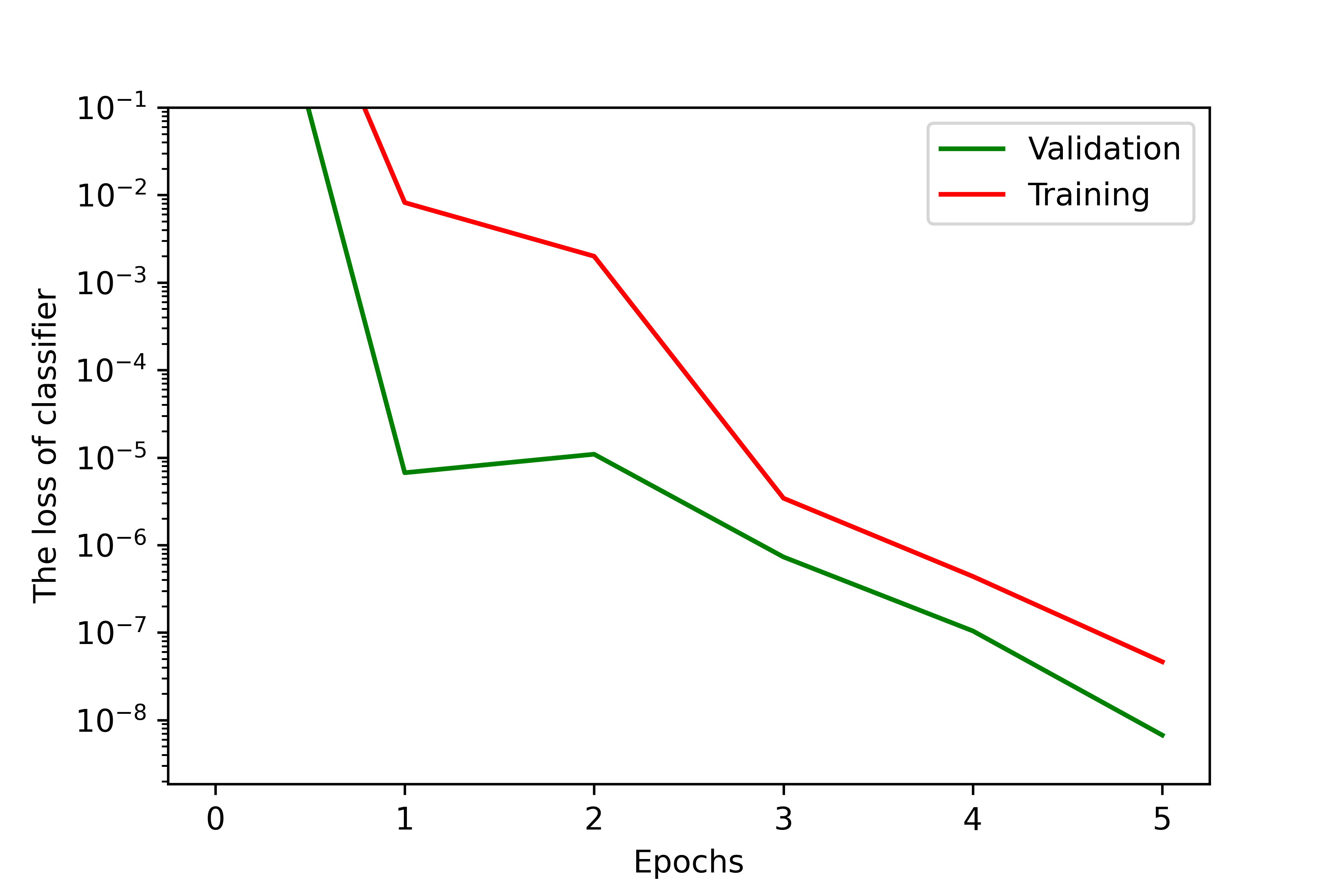}
\caption{Training and validation losses of classifier.}\label{fig:Classi}
\end{figure}

Figure~\ref{fig:Classi} shows the training and validation loss of classifier. 
One can find that the training and validation losses reach $10^{-7}$ and $10^{-8}$ at epoch = 5, respectively. The accuracy is defined as $\alpha =\sum_{i} \bm{M}_{i,i}/ \sum_{i,j} \bm{M}_{i,j}$, where $\bm{M}$ is the confusion matrix \cite{hagita2021deep}, a visual table layout that reflects the predictions of the network. As shown in Table~\ref{Tab:3},  $\bm{M}_{ij}$ denotes the number of $i$ identified to be $j$, and $i, j\in\{\text{LAM, HEX}\}$. These results show that the classifier can 100\% identify structures.

\begin{table}[!htbp] 
\centering
\caption{The confusion matrix of classifier at epoch = 5. \label{Tab:3}}
\begin{tabular}{|c|c|c|}
\hline
 & Identified LAM & Identified HEX \\
\hline Certain LAM & 13100 & 0 \\
\hline Certain HEX &0 & 13500  \\
\hline
\end{tabular}
\end{table}

We classify the training and validation data by the classifier, and obtain 188100 LAM and 160000 HEX phases. Then we adopt them as the training and validation data for each subnet. For the test data onto each subnet, we only consider the data translated by $\bm{t} = (0.17,0.17)^{\top}$ in the test set. Table \ref{Tab:split1} indicates the size of dataset for each SPM subnet.
\begin{table}[!htbp]
\centering
\caption{Amount of training, validation and test data for LAM and HEX subnets.\label{Tab:split1}} 
\begin{tabular}{cccc}
\toprule[1pt] 
  & \text{Training data} & \text{Validation data} & \text{Test data} \\ 
\hline\text { LAM subnet }  & 150480 & 37620 & 2620 \\
\text { HEX subnet }  & 128000 & 32000 & 2700 \\
\bottomrule[1pt]
\end{tabular}  
\end{table}

\begin{figure}[!htbp]
  \centering
  \subfigure[LAM subnet]{
  \centering
      \includegraphics[width=7cm]{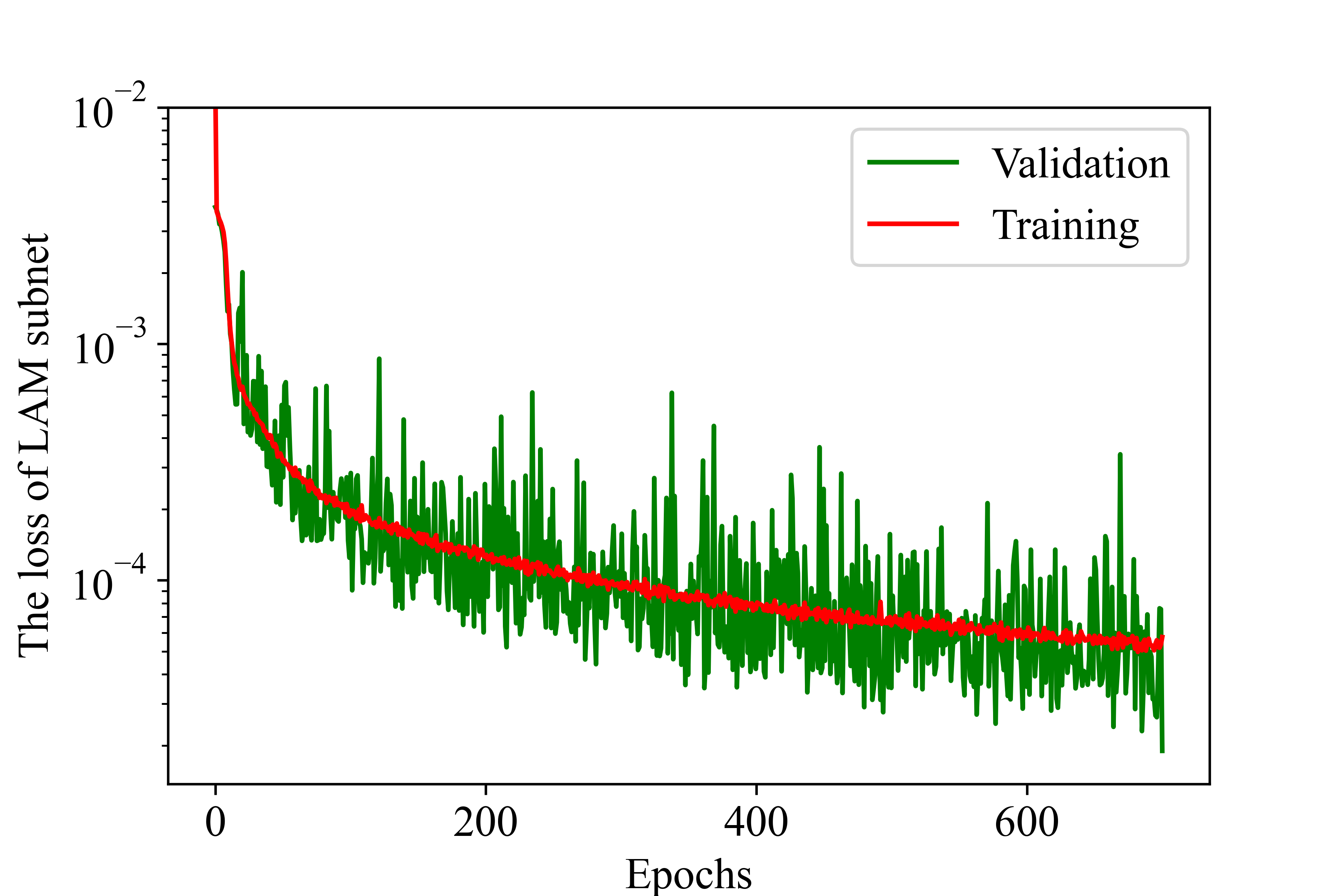}
  }
  \subfigure[HEX subnet]{
  \centering
    \includegraphics[width=7cm]{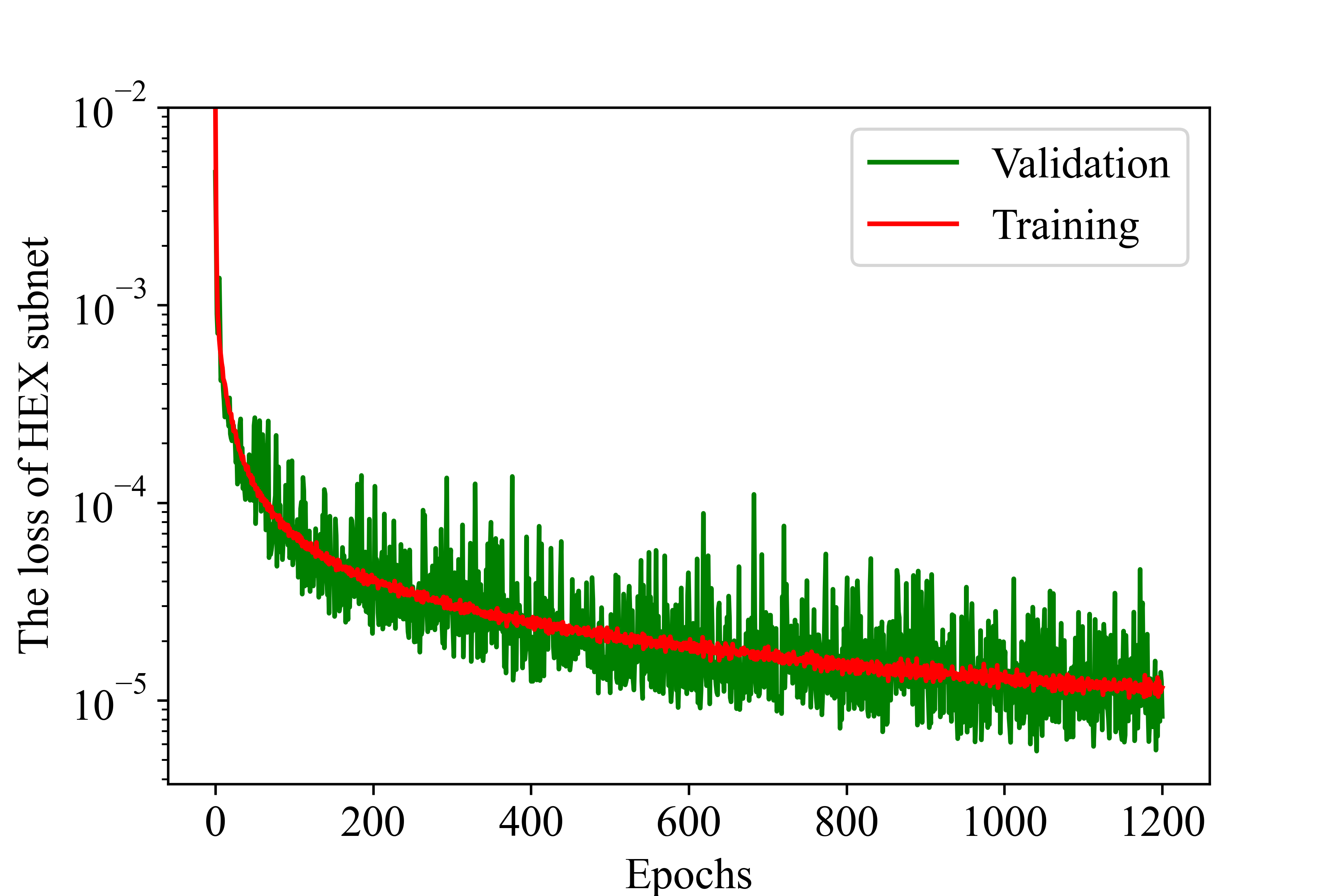}
  }
  \caption{The training and validation losses of SPM subnets: (a) LAM and (b) HEX.}
  \label{Fig:sub}
\end{figure}

Figure~\ref{Fig:sub} presents the training and validation losses of SPM networks. We can see that, the training loss of the LAM (HEX) subnet is $5.75 \times10 ^ {- 5} $ ($1.16 \times10 ^ {- 5} $), the validation loss is $1.89 \times10 ^ {- 5} $ ($8.30 \times10 ^ {- 6} $) at epoch = 700 (1200).  

\begin{figure}[!htbp]
  \centering
  \includegraphics[width=12cm]{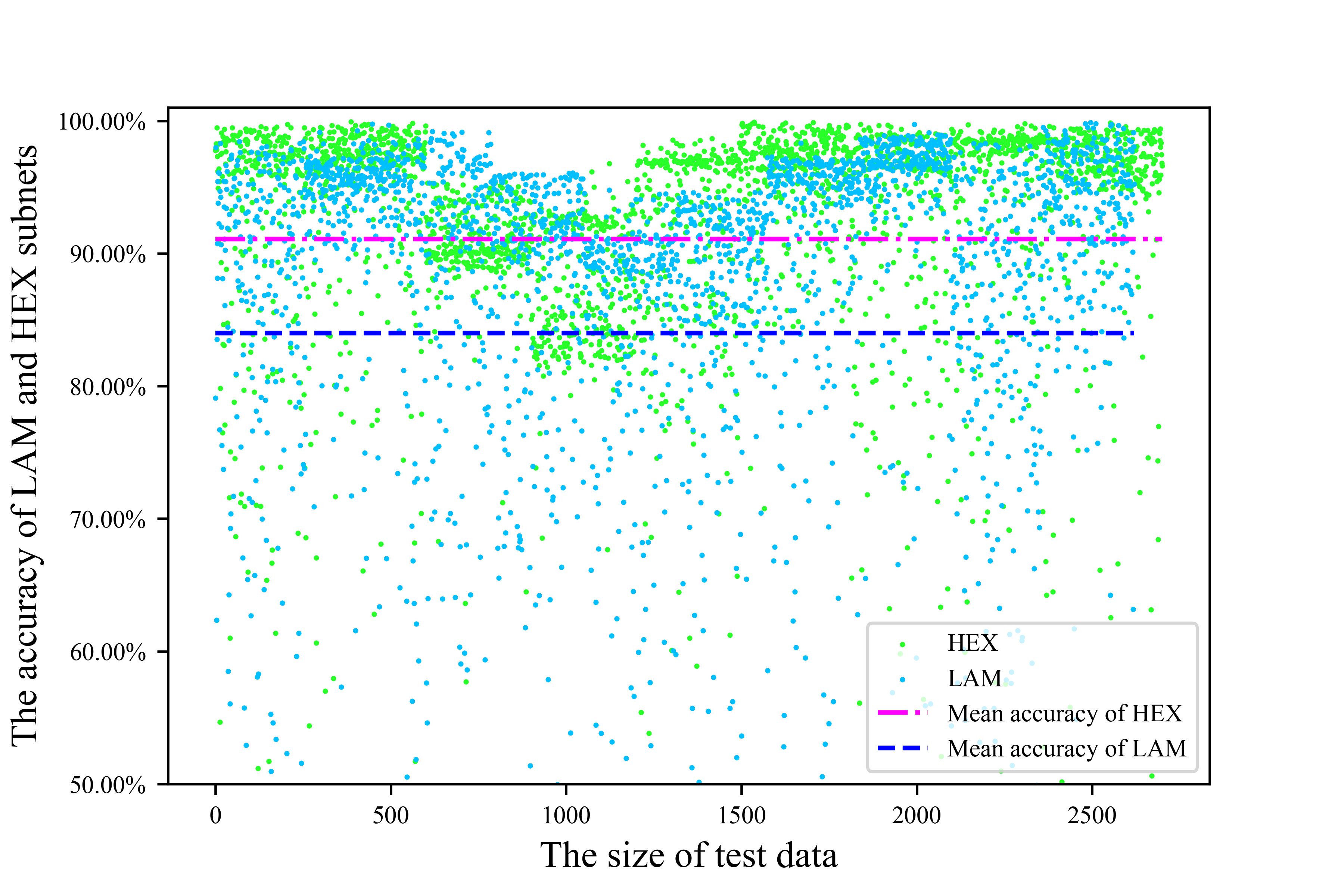}
\caption{The test accuracy of LAM (blue) and HEX (green) subnets. Scatter plots demonstrate the parameters prediction accuracy of HEX and LAM subnets. The average test accuracy of the two subnets is shown by two dashed lines.  \label{Fig:test}}
\end{figure}

\begin{figure}[htbp]
\centering
  \subfigure[Targeted LAM]{
  \centering
    \includegraphics[width=3.8cm]{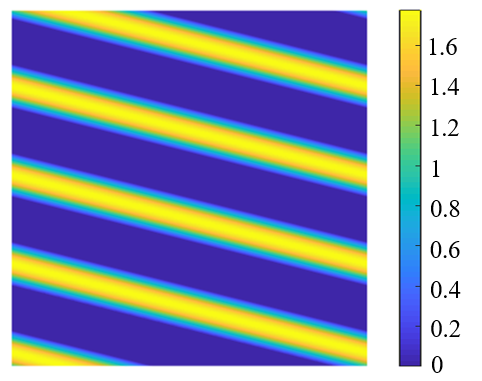}
  }
  \quad
  \subfigure[Predicted LAM]{
  \centering
    \includegraphics[width=3.8cm]{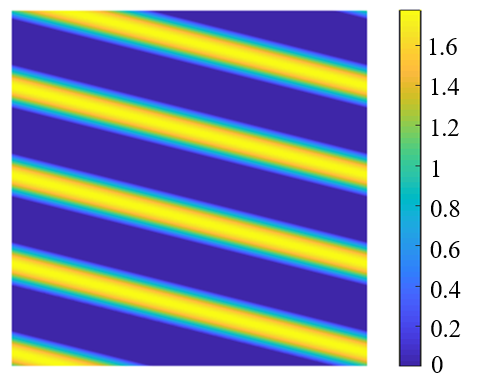}
  }
  \quad
  \subfigure[Absolute error]{
  \centering
    \includegraphics[width=4.2cm]{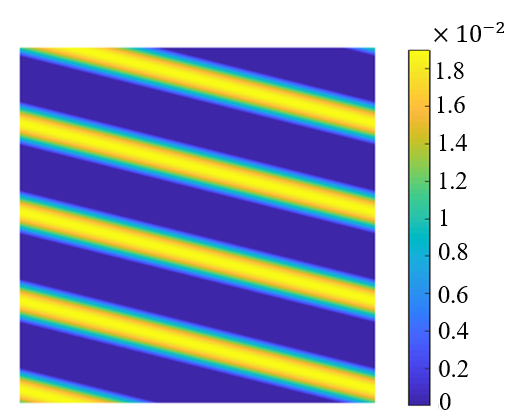}
  }\\
  \subfigure[Targeted HEX]{
  \centering
    \includegraphics[width=4.35cm]{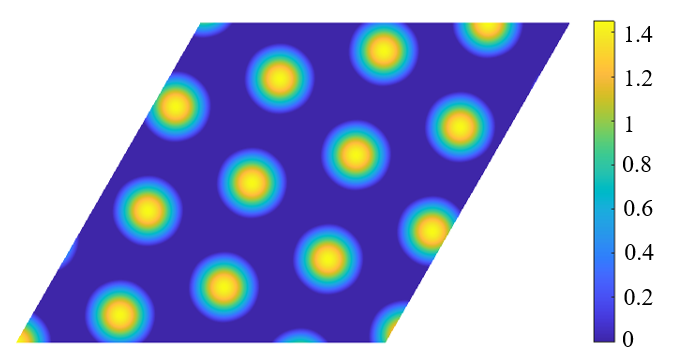}
  }
  \subfigure[Predicted HEX]{
  \centering
    \includegraphics[width=4.35cm]{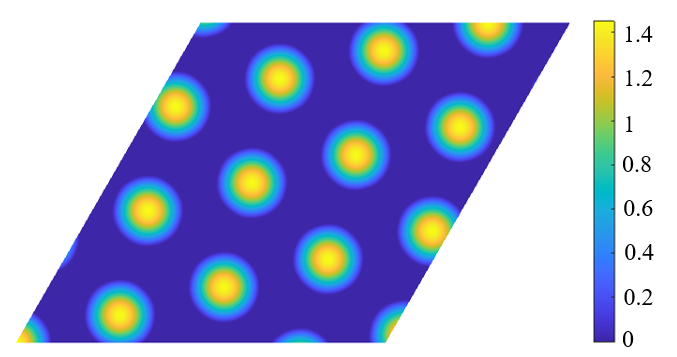}
  }
  \subfigure[Absolute error]{
  \centering
    \includegraphics[width=4.45cm]{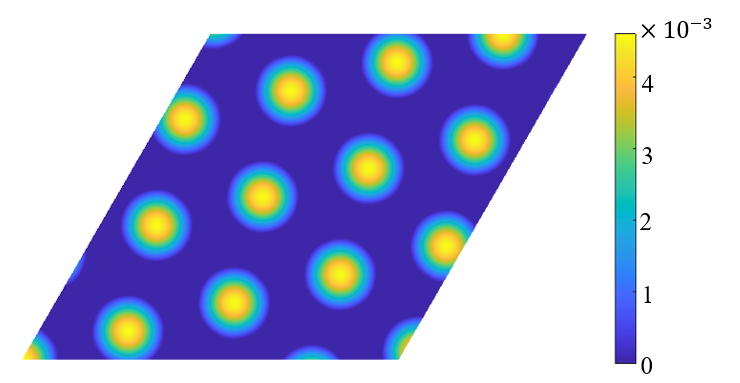}
  }  
  \caption{The phases with targeted and predicted parameters are obtained by solving direct problem (\ref{eq:Min}). LAM phase: (a) [$\tau^{*}$,$\gamma^{*}$] = [-0.4, 0.2], $\theta = 82\degree$ and $\boldsymbol{t} = (0.17,0.17)^{\top}$; (b) [$\tau$,$\gamma$] = [-0.3914, 0.2067], $\theta = 82\degree$ and $\boldsymbol{t} = (0.17,0.17)^{\top}$. HEX phase: (d) [$\tau^{*}$,$\gamma^{*}$] = [-0.08, 0.28], $\theta = 40\degree$ and $\boldsymbol{t} = (0.17,0.17)^{\top}$; (e) [$\tau$,$\gamma$] = [-0.0807, 0.2744], $\theta = 40\degree$ and $\boldsymbol{t} = (0.17,0.17)^{\top}$. Absolute errors between targeted and predicted LAM (c) and HEX (f) phases.
  \label{Fig:lamhex}}
\end{figure}

 We define the relative error

$$
\mathcal{E}=\frac{\left\|\mathbf{u}-\tilde{\mathbf{u}}\right\|_{\ell^2}}{\left\|\mathbf{u}\right\|_{\ell^2}},
$$
and the average relative error 
$$
\mathcal{E}_{\text {average }}=\frac{1}{N_{test}}\sum_{k=1}^{N_{\text {test}}} \frac{\left\|\mathbf{u}^k-\tilde{\mathbf{u}}^k\right\|_{\ell^2}}{\left\|\mathbf{u}^k\right\|_{\ell^2}},
$$
where $N_{\text {test}}$ is the size of test data. The predicted accuracy of single sample is $(1-\mathcal{E}) \times 100 \%$, while the average accuracy is $(1-\mathcal{E}_{\text {average}}) \times 100 \%$.  Figure~\ref{Fig:test} illustrates the test accuracy of the LAM and HEX subnets. Results indicate that the accuracy of parameters prediction for a single sample can achieve $80\%\sim 100\%$, and the average prediction accuracy of LAM subnet (blue dashed line) is 84\%. For the HEX subnet (pink dashed line), the average prediction accuracy can reach 91\%.
We select randomly two test samples in the test set and input them into the network to predict the corresponding parameters. Next, we obtain structures corresponding to the predicted parameters by solving the direct problem. Figure~\ref{Fig:lamhex} shows LAM and HEX phases with targeted and predicted parameters, and the absolute errors between the targeted and predicted phases. From Figure~\ref{Fig:lamhex}\,(c)(f), we can see that the absolute error 
of LAM phase is $10^{-2}$, and HEX phase is $10^{-3}$. This also reflects a good fitting effect of SPM subnets.

\newpage
\section{Conclusion}
\label{sec:conclusion}

In this paper, we propose a multi-category neural network for inverse design of ordered periodic structures. The proposed network can construct the mapping between phases and physical parameters. For periodic phases, we give an extensible RSDA approach to augment data. Then we apply these methods to two-dimensional diblock copolymer system. The dataset is produced by LB free energy functional.
Experimental results show that the structure recognition accuracy of 
the classifier can reach 100\% based on 26600 randomly selected test data. Moreover, on a dataset consisting of 5320 randomly selected test data, the parameters prediction accuracy of the LAM phase reaches 84\%, and the accuracy of the HEX phase reaches 91\%. The network model and RSDA method are applied to a two dimensional problem, however, they can be extended to higher-dimensional inverse design problems.

	
\bibliographystyle{unsrt}
\bibliography{RSbib}	
\end{document}